\documentclass[reprint,...]{revtex4-1}

\usepackage{graphicx}

\usepackage{amsmath}
\usepackage{amsfonts}
\usepackage{amssymb}

\usepackage{appendix}
\usepackage{fancyvrb}
\usepackage{gensymb}
\usepackage{verbatim}
\usepackage{wrapfig}
\usepackage{bm}
\usepackage{url}
\usepackage{xfrac}
\usepackage{xcolor}
\usepackage[english]{babel}
\usepackage{setspace}
\usepackage{csquotes}

\usepackage{array}
\newcolumntype{L}[1]{>{\raggedright\let\newline\\\arraybackslash\hspace{0pt}}m{#1}}
\newcolumntype{C}[1]{>{\centering\let\newline\\\arraybackslash\hspace{0pt}}m{#1}}
\newcolumntype{R}[1]{>{\raggedleft\let\newline\\\arraybackslash\hspace{0pt}}m{#1}}

\newcommand{\abs}[1]{\left| #1 \right|} 
\newcommand{\pd}[2]{\frac{\partial #1}{\partial #2}} 
 
\let\baraccent=\= 
\renewcommand{\=}[1]{\stackrel{#1}{=}} 

\usepackage{bm}

\begin{document}
	
	
	\title{Generation of synthetic magnetized plasma turbulence by representation as a stochastic process}
	
	
	\author{J.Leddy}
	\email{jleddy@txcorp.com}
	\affiliation{Tech-X Corporation, Boulder CO, USA}
	\affiliation{York Plasma Institute, University of York, Heslington UK}
	\author{C.Bowman}
	\author{K.Gibson}
	\author{B.Dudson}
	\affiliation{York Plasma Institute, University of York, Heslington UK}
	
	
	\date{\today}
	
	\begin{abstract}
		Plasma turbulence simulations are often computationally expensive with delicate numerical stability.  Yet, long simulations are needed to generate uncorrelated turbulence data for studies such as microwave scattering through density perturbations.  For this reason, alternative methods of producing accurate synthetic turbulence profiles via statistical methods is of interest. Such a method is proposed where the two-point covariance function of the desired turbulence is used to construct a multi-variate normal distribution. Sampling from this distribution produces random fields which are both qualitatively and quantitatively similar to the input turbulence data set. The resulting `synthetic' turbulent profiles are uncorrelated in `time' so it is useful only for scenarios that do not require such physical evolution.
	\end{abstract}
	
	
	\maketitle

\section{Introduction}
Turbulence is a ubiquitous fluid phenomenon that has often significant and unexpected effects on transport of energy and density within the fluid as well as the propagation of waves throughout it.  Due to the analytically intractable nature of the fluid equations, numerical modeling is often the only reliable method for predicting and understanding the fluid behavior. These simulations can be very computationally expensive, depending on the method used.  Kinetic simulations are possible, but only for very small scale lengths, which often do not capture the required diversity of scale intrinsic to turbulence.  For this reason, fluid codes are often the tool of choice, depending on the physics of interest. In cases where the turbulent evolution is unimportant, and instead uncorrelated snapshots of the turbulent density/temperature structure are required (such as electromagnetic propagation through plasma turbulence \cite{Eliasson2016,Kohn2016} or statistical transport of energy and particles in the tokamak scrape-off-layer \cite{Mekkaoui2012}), fluid simulations are expensive. Instead, methods for generating these snapshots without the use of turbulence simulations are preferred.

In this paper we present such a method, which can generate realistic synthetic turbulence via direct sampling of a specially constructed multivariate normal distribution. Other methods exist for generating synthetic turbulence for such purposes, including inversion of Fourier spectra with imposed stochastic noise, and summation of Gaussian distributions with normal stochastic spacing \cite{Baran1995,Aubard2012}.  Though these methods can reproduce spatial spectra accurately, they are not guaranteed to obtain the correct spatial covariance.  The method presented herein  accurately reproduces both of these properties, as will be discussed in later sections.
\subsection{Method}
\begin{figure}
	\centering
	\includegraphics[width=\linewidth]{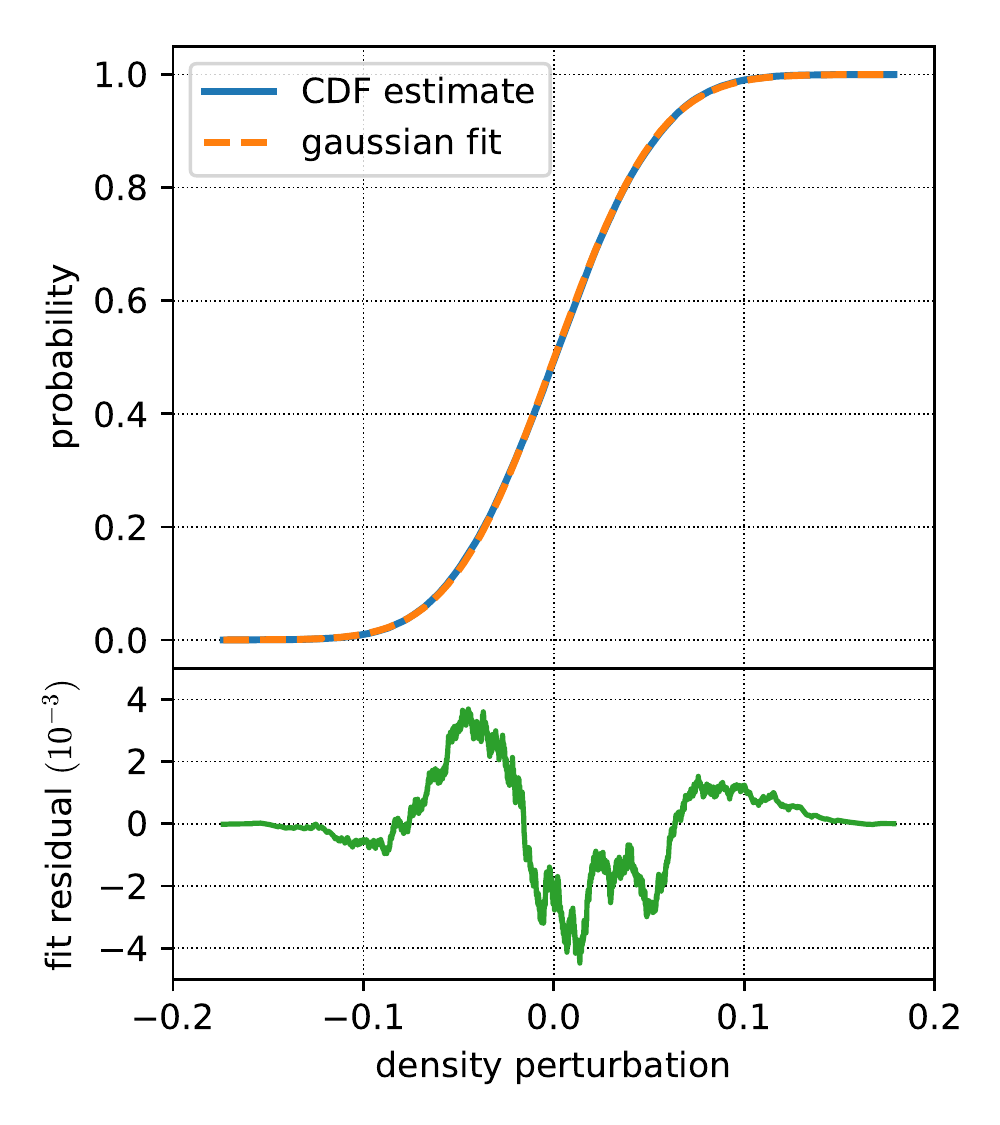}
	\caption{The distribution of densities for a given point in turbulence data generated by the Hasegawa-Wakatani model is very well represented by a Gaussian distribution (top) with residuals only as large as $0.004$ (bottom).}
	\label{fig:MVN_justification}
\end{figure}

The stochastic nature of magnetized plasma turbulence dictates that, under steady-state conditions, the distribution of values of an associated field (e.g. the density) at a particular point over a long period of time will converge to some stationary distribution. In some cases, such as the Hasegawa-Wakatani model \cite{Wakatani1984} (which is subsequently discussed in further detail) this stationary distribution is Gaussian, as shown in figure \ref{fig:MVN_justification}. In this article we describe how plasma turbulence with this property may be effectively modeled using \emph{Gaussian processes}.  Though a normal distribution is used herein because it describes the model data, in principle any appropriate analytical distribution function can be used to fit the desired stationary distribution and be used for the random sampling.

A Gaussian process is a statistical model in which the value of a field at any point in some continuous space of interest is normally distributed, and the joint distribution of any finite collection of these points is multivariate normal. Additionally, the covariance between the values of the field at any two points in the space $\underline{x}_i$ and $\underline{x}_j$ is defined by some covariance function $K(\underline{x}_i , \underline{x}_j)$ which depends only on the spatial location of the two points.
If $\mathbf{y} = [y_1, y_2, \dots , y_N]$ are the values of the field at some collection of points $\{\underline{x}_i\}_{1\le i \le N}$, then
\begin{equation}
\mathbf{y} \sim \mathcal{N}\left( \bm{\mu}, \mathbf{\Sigma} \right),
\end{equation}
where
\begin{equation}
\Sigma_{ij} = K(\underline{x}_i , \underline{x}_j).
\end{equation}

The mean vector $\bm{\mu}$ may be chosen to represent the background profile of the field, or simply taken to be zero in the case of perturbed quantities. The stochastic properties of the model are therefore entirely dictated by the covariance matrix $\mathbf{\Sigma}$ and by extension the covariance function $K(\underline{x}_i , \underline{x}_j)$. In order to model plasma turbulence fields through Gaussian processes, we must therefore characterize the spatial covariance function of whichever type turbulence is desired - this procedure is discussed in the following section.

\begin{figure}
	\centering
	\includegraphics[width=\linewidth]{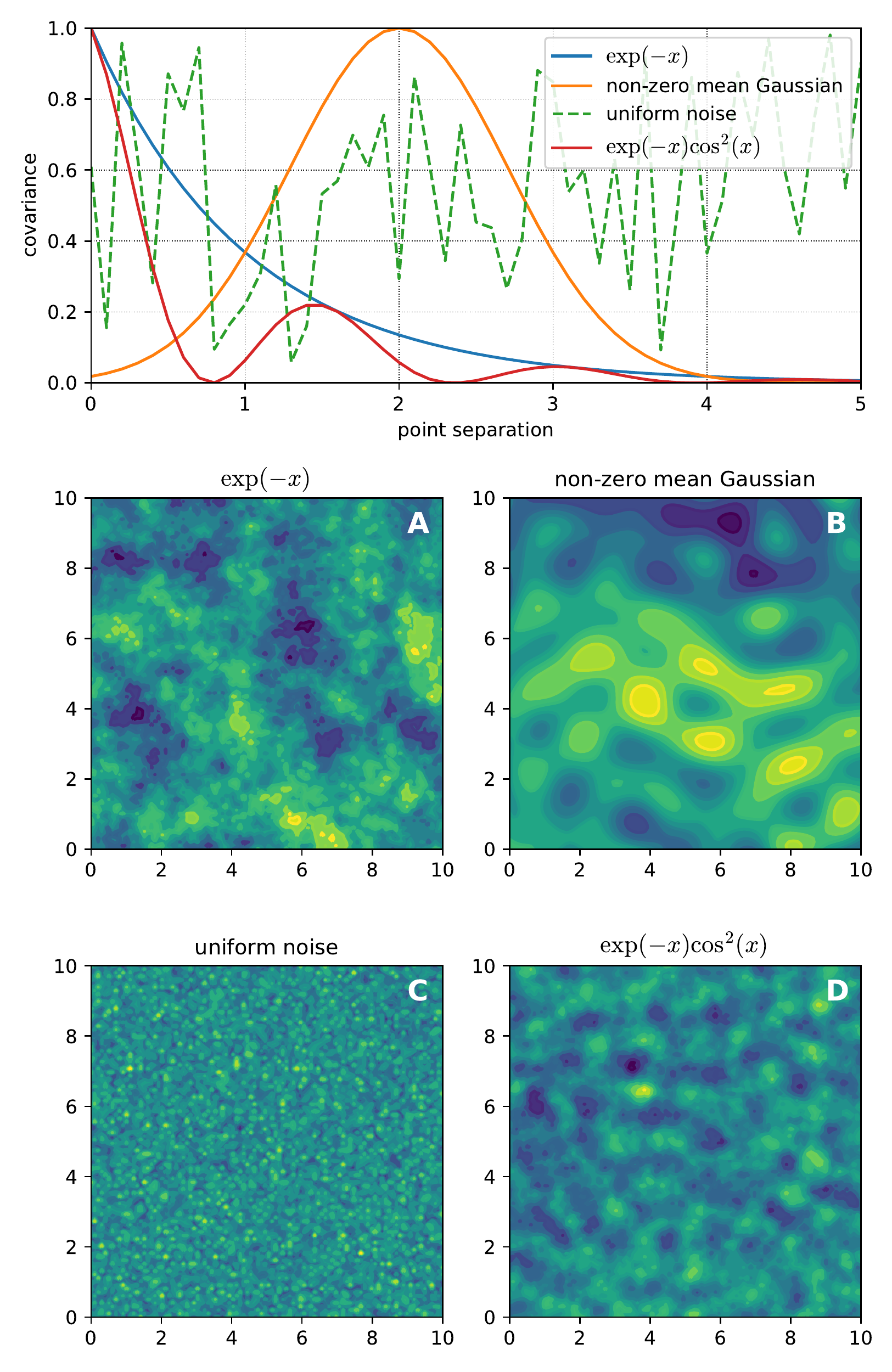}
	\caption{Four example covariance functions are shown (top), each showing unique features in the resulting field sample (A-D).}
	\label{fig:multiple_covariances}
\end{figure}

Once the covariance function is known, the joint distribution for a chosen set of points in the space may be constructed, and random fields generated simply by drawing samples from this distribution. Examples of samples drawn from Gaussian processes with different covariance functions are shown in figure \ref{fig:multiple_covariances}, and illustrate how the covariance function dictates the spatial structure of the generated fields.  This figure is discussed in more detail in section \ref{sec:other_applications}.

In order to efficiently draw samples from the joint distribution, we apply the Karhunen-Lo\`eve transform, which in the special case of a multivariate normal distribution allows the sample to be expressed as a linear combination of independent normal random variables. The first step is to compute the eigen-decomposition of the covariance matrix $\mathbf{\Sigma} = \mathbf{Q \Lambda Q}^{-1}$ where
\begin{equation}
	\mathbf{Q} = 
	\begin{bmatrix}
	| & | &  & | \\
	\mathbf{v}_1 & \mathbf{v}_2 & \hdots & \mathbf{v}_m \\
	| & | &  & |
	\end{bmatrix}
	, \quad
	\mathbf{\Lambda} = 
	\begin{bmatrix}
	\lambda_1 &  &  &  \\
	& \lambda_2 &  &  & \\
	&  & \ddots &  & \\
	&  &  & \lambda_m &
	\end{bmatrix}
\end{equation}
and $\mathbf{v}_i$, $\lambda_i$ are the $i$'th eigenvector and eigenvalue of $\mathbf{\Sigma}$ respectively. The field values $\mathbf{y}$ can now be expressed as
\begin{equation}
\mathbf{y} = \bm{\mu} + \mathbf{Q u},
\end{equation}
where
\begin{equation}
\mathbf{u} \sim \mathcal{N}\left( 0, \mathbf{\Lambda} \right).
\end{equation}
New samples can now be easily produced by drawing a new $\mathbf{u}$, which is computationally inexpensive because $\mathbf{\Lambda}$ is diagonal.
\section{Covariance function estimation}
\begin{figure}
	\centering
	\includegraphics[width=\linewidth]{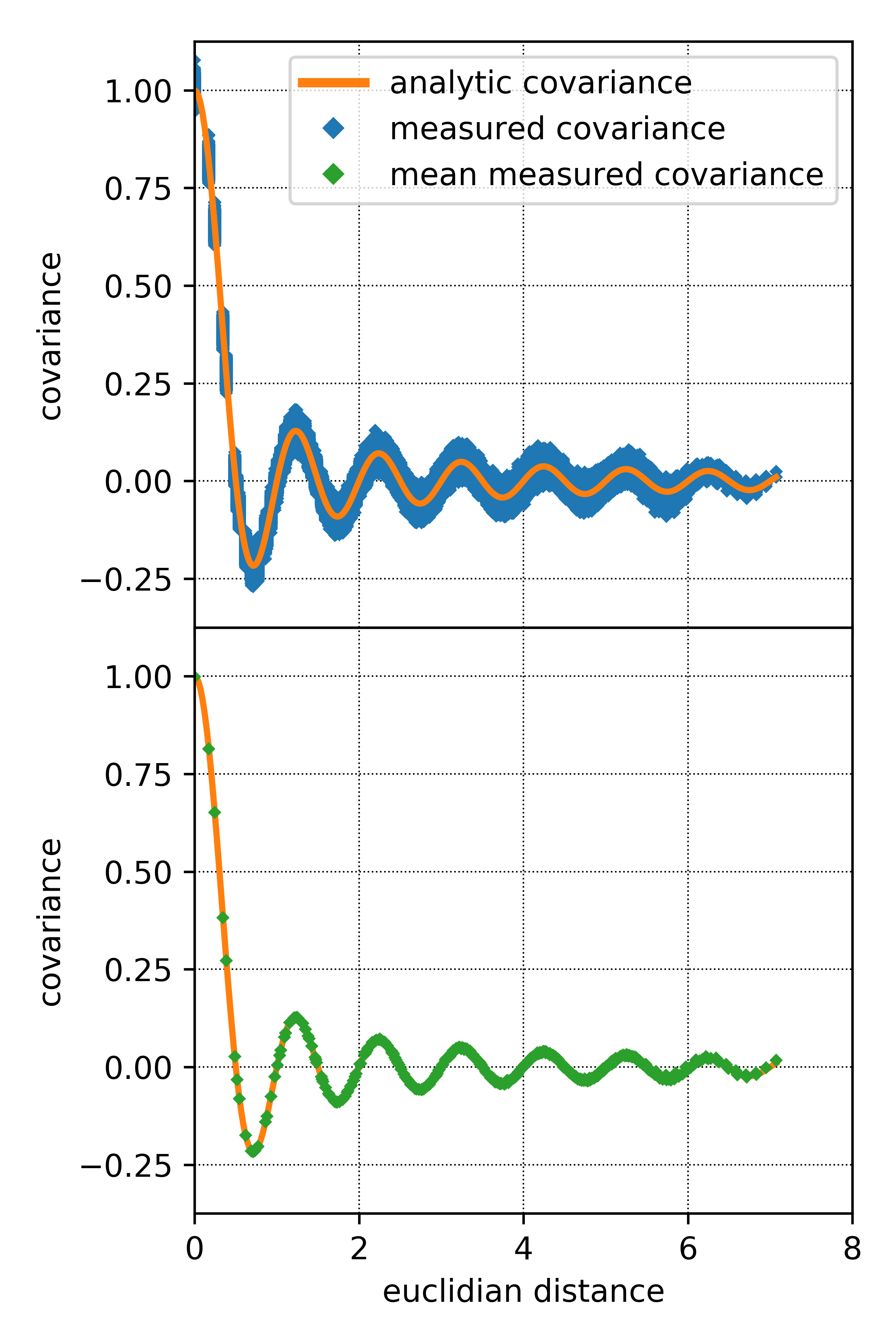}
	\caption{A given value of point separation has a distribution of observed covariance values for finitely many samples (top), and the mean of this distribution is an effective estimator for the true covariance (bottom).}
	\label{fig:covariance_validation}
\end{figure}

To ensure that we are able to correctly infer a spatial covariance function from turbulence simulation data, a dataset with known covariance was analysed. This testing dataset consisted of 5000 images and was produced via the Gaussian process sampling approach described in the previous section, using the following covariance function $K(\underline{x}_i , \underline{x}_j) = \mathrm{sinc}\,\left(\sqrt{(\underline{x}_i - \underline{x}_j)^{\top}(\underline{x}_i - \underline{x}_j)} \right)$. Results obtained from reconstructing the covariance function of the testing dataset are shown in figure \ref{fig:covariance_validation}.

Any covariance calculated using a finite set of samples is merely an estimator of the underlying true covariance (only being exact in the limit of infinitely many samples) - this is reflected in spread of observed covariances at a given separation. This spread can be significantly reduced however, by averaging over covariance estimates which share the same separation, as shown in the lower plot of figure \ref{fig:covariance_validation}. This procedure recovers the original covariance function with impressive accuracy, and is used in the following section to estimate the covariance function of turbulence data generated using the Hasegawa-Wakatani model.

\section{Generation of synthetic turbulence profiles}
To reproduce turbulence with the desired traits, the covariance function of the turbulence must be known.  This can be obtained through experimental measurements or simulation.  Using a simulation to obtain the covariance function requires a relatively short simulation to turbulent saturation only - this can then be used to efficiently generate unlimited amounts of uncorrelated turbulence profiles with the same features.  As a proof of concept, a simple two-dimensional plasma turbulence model has been chosen here for generation of these initial profiles.
\subsection{Turbulence simulation covariance}
The Hasegawa-Wakatani model \cite{Wakatani1984} is a two-dimensional resistive drift-wave turbulence model relevant to magnetized plasmas.  The equations evolve the density, $n$, and vorticity, $\varpi$:
\begin{eqnarray}
\pd{n}{t} &=& -[ \phi,n ] - \kappa \pd{\phi}{z} - \alpha(\phi-n) \\
\pd{\varpi}{t} &=& -[ \phi,\varpi ] - \alpha(\phi-n)
\end{eqnarray}
where $\kappa$ represents a background density gradient in the $x$-direction, $\alpha$ is proportional to the parallel conductivity, $\phi=\nabla^{-2}\varpi$ is the electric potential calculated by inverting the vorticity, and the bracket operator is defined such that $[f,g] = \pd{f}{x}\pd{g}{z}-\pd{f}{z}\pd{g}{x}$.  The background magnetic field is in the $y$-direction, so the $x-z$ plane simulated is the perpendicular plane.

\begin{figure}
	\centering
	\includegraphics[width=\linewidth]{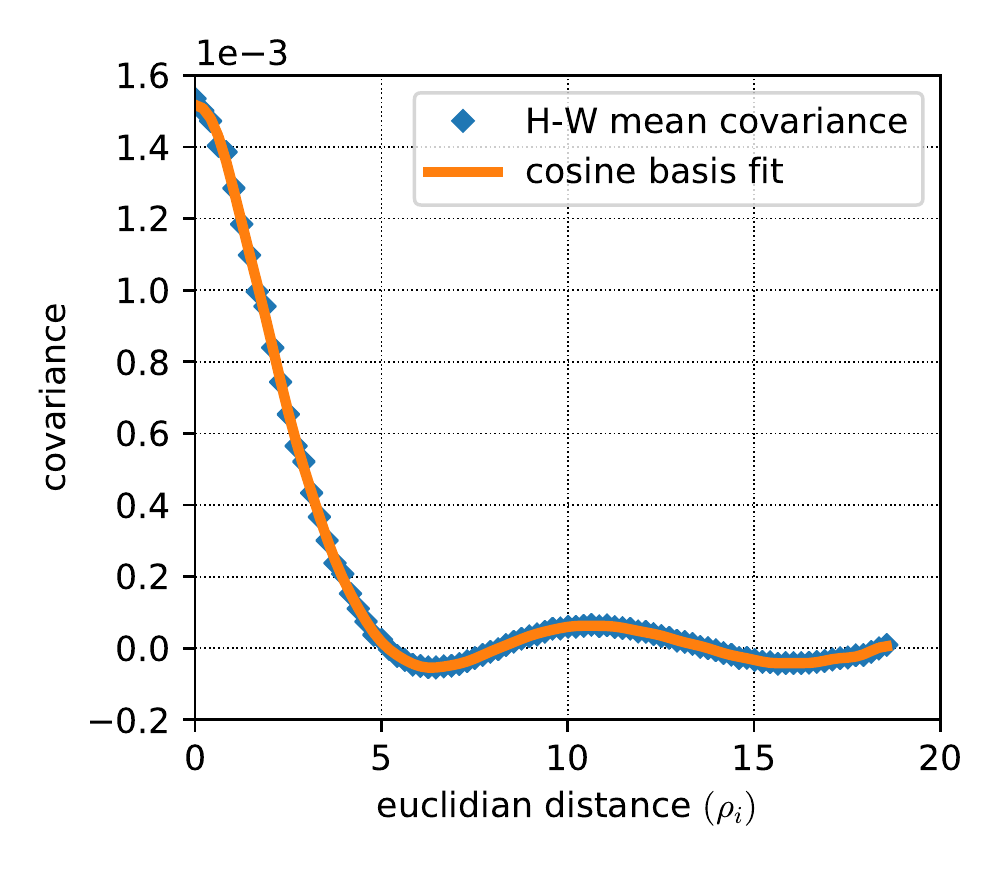}
	\caption{The estimated covariance function of the Hasegawa-Wakatani turbulence is shown (points), and is well-modelled using a sum of cosine basis functions (line) as in equation \ref{eqn:cosines}.}
	\label{fig:HW_cov}
\end{figure}

The BOUT++ framework \cite{Dudson2009} was used to evolve these equations to a saturated state producing turbulence with an estimated covariance function seen in figure \ref{fig:HW_cov}. The covariance function was parameterised through representation as a sum of cosine basis functions such that on some interval $[0,L]$
\begin{equation} \label{eqn:cosines}
K(\underline{x}_i , \underline{x}_j) \approx \sum_{n=0}^{N} a_n \cos{\left(\frac{n\pi}{L}\sqrt{(\underline{x}_i - \underline{x}_j)^{\top}(\underline{x}_i - \underline{x}_j)} \right) }.
\end{equation}

\section{Validation of synthetic turbulence}
\begin{figure*}[t]
	\centering
	\includegraphics[width=\linewidth]{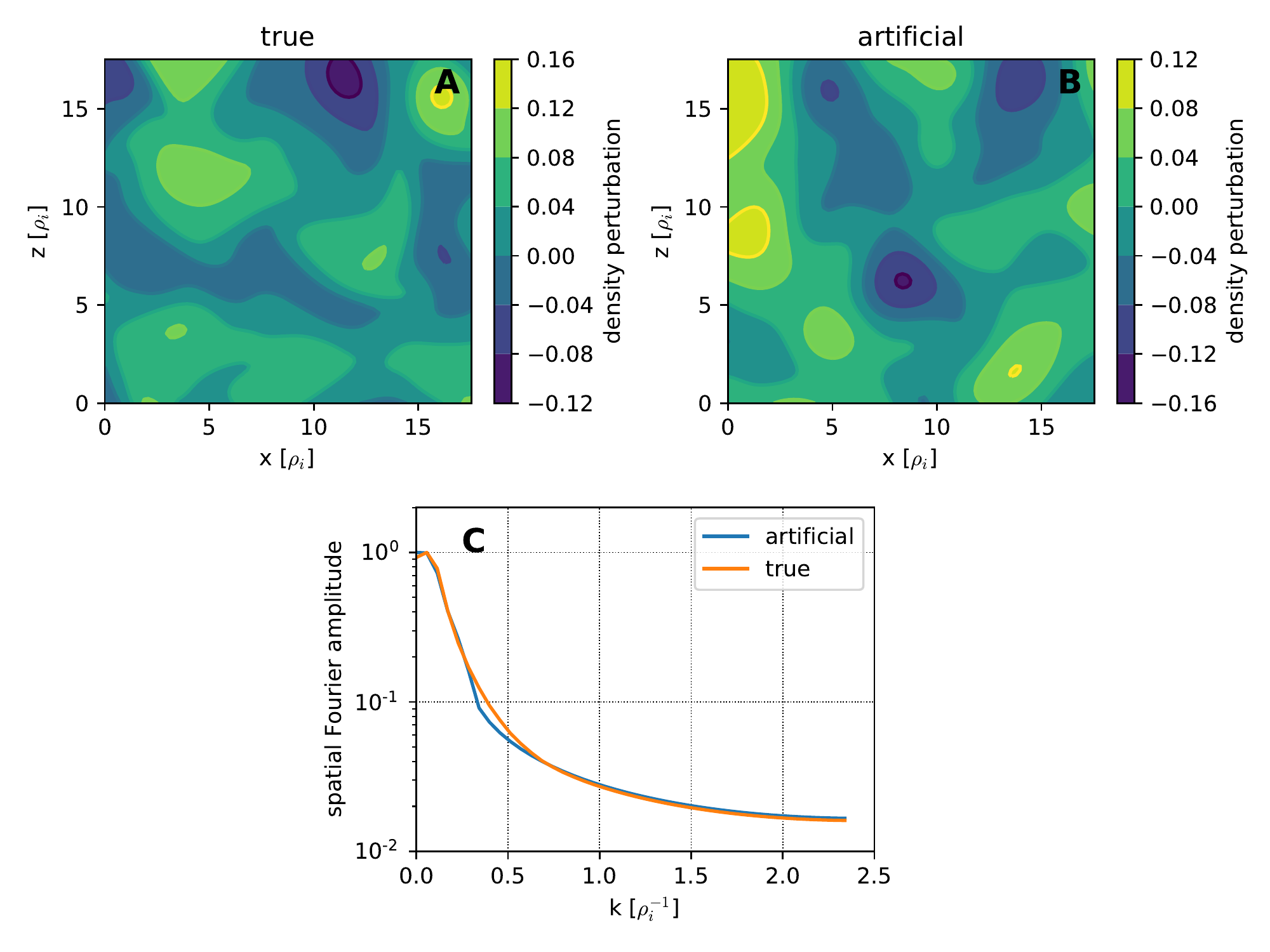}
	\caption{The true (A) and synthetic (B) turbulence are qualitatively similar.  The spatial spectra (C) agree at small and large wave number, and differ by roughly $25\%$ at mid wave number.  The magnitude in the differing regions, however, is relatively low so that this difference is not significant.}
	\label{fig:comparison}
\end{figure*}

The turbulence produced via sampling from the multivariate normal distribution is shown in comparison to the original Hasegawa-Wakatani turbulence in figure \ref{fig:comparison}.  By design, the two-point covariance function is on average the same as that of the simulated turbulence.  Another feature of the turbulence is the spatial spectrum, which shows the relative amplitude of the density perturbation signal that exists at each wave number.  Qualitatively, the synthetic and original turbulence are very similar, agreeing at large and small inverse structure size.  There is agreement within 5\% everywhere except for a peak of 25\% error around 0.4$\rho_i^{-1}$, however the absolute difference is relatively small at this point since the spectral magnitude is a tenth of the maximum value.

Theoretically this agreement is expected because the distribution of spatial scales are inherently captured by the covariance function.  Assuming the turbulence is a Gaussian process, which figure \ref{fig:MVN_justification} supports, then by definition the correct covariance will provide a matching spectrum as the processes are identical.  Any differences between the spectra can therefore be attributed to one or both of the following: The Hasegawa-Wakatani turbulence cannot be represented exactly as a Gaussian process, or our assumption that the covariance is spatially invariant is an over-simplification, and results in an estimated covariance function which cannot fully reproduce the desired behaviour.

\subsection{Limitations}
The primary limitation of this process to create synthetic turbulence is that the generated turbulent structures will always be consistent throughout the domain because the covariance function must be stationary and isotropic.  This means complex features with strong local correlations, such as single shocks or swirls are unable to be reproduced due to their local anisotropic covariance.  In cases where many samples are required for statistical averaging, such as with full-wave electromagnetic propagation through plasma turbulence, it is thought that these features will average out, but this is an important consideration to be made on an individual basis for each application.

Another limitation of the method is that the generated profiles will always have an isotropic distribution of turbulence amplitude across the domain, and will therefore not reproduce trends in background or fluctuation profiles.  However, turbulent fluctuations can be scaled after generation to match experimental or simulation profile envelopes.

\subsection{Other applications} \label{sec:other_applications}
A Gaussian process is completely specified by the choice of covariance function, and depending on the particular function used the resulting fields can possess interesting or unique features - this is illustrated by figure \ref{fig:multiple_covariances} where four example random fields are plotted alongside the covariance functions used to produce them. For the two covariance functions containing the exponential decay, a non-zero gradient exists at zero separation, meaning that there is a non-zero difference in covariances for two very close points.  This results in the small scale structures seen in the resulting fields for these two covariances.  The Gaussian covariance function is not centered about zero, indicating that the strongest correlation exists between points that are separated by a distance of 2 (arbitrary units).  This naturally leads to a periodicity in the resulting field that has a wavelength of 2, but the overall structure is still irregular.  The field resulting from sampling the distribution with a uniformly random covariance function appears to have no distinct features.  Upon examination of its spatial spectrum it was found to be white noise (ie. the spatial spectrum power is uniformly distributed across all wave numbers).  This large variety of structures and features that can be generated shows the flexibility of this method and expands the possible applications.

Each sample of the normal distribution will produce a unique slice of turbulence uncorrelated with the other samples.  However, it may be possible to parameterize the temporal correlation to include as an extra dimensionality in the covariance matrix, approximating the evolution of the turbulence.  This evolution, however, would not necessarily be physical, though it would demonstrate the correct temporal correlation.  For applications that require temporal accuracy, direct fluid/kinetic simulation is more appropriate.  Other methods, such as the synthetic-eddy-method, are able to produce uncorrelated snapshots like the method presented here but with a physical representation of the flows \cite{Pinon2017,Mekkaoui2013}.  These methods trade efficiency for accuracy, appealing to a different set of applications.

Finally, an generalisation of this method could be used to generate turbulence with a spatially varying non-stationary covariance \cite{Risser2015}.  This would allow for the generation of samples that contain different turbulent properties throughout the domain (eg. turbulent structures that vary as you move across the separatrix in a tokamak).    Relaxing the assumption that $K(\underline{x}_i , \underline{x}_j) = K(\abs{\underline{x}_i - \underline{x}_j})$ would allow such a scenario, but would also increase the dimensionality of the covariance function thus significantly increasing the computational cost of generating the distribution from which samples are taken.

\section{Conclusion}
By treating turbulence as a Gaussian process, a covariance function can be used to generate a multivariate normal distribution from which synthetic turbulent structures may be sampled.  These turbulent profiles will have, on average, the same spatial covariance as that used to generate the normal distribution.  Assuming this covariance function is taken from a measurement or simulation of a turbulent system, the synthetic turbulence will have the same atemporal properties of the original system.  This method also ensures that the spatial spectrum of the synthetic turbulence also matches that of the original system.  This method can be, as envisioned here, used to efficiently generate turbulent profiles for the investigation of the propagation of electromagnetic waves through plasma turbulence.  Other possibilities include statistical analysis of transport and flux and the production of turbulent structures for animation/CGI.  Presented here is a tool to efficiently support other areas of research.

\subsubsection*{Acknowledgements}
This work has received funding from the RCUK Energy Programme, grant number EP/I501045.  It was also received support from EPSRC grant EP/M001423/1.  It has been carried out within the framework of the EUROfusion Consortium and has received funding from the Euratom research and training programme 2014-2018 under grant agreement No 633053.  The views and opinions expressed herein do not necessarily reflect those of the European Commission.

\bibliographystyle{unsrt} 
\bibliography{master}

\end{document}